# Large-scale earthquake sequence simulations on 3D nonplanar faults using the boundary element method accelerated by lattice H-matrices

So Ozawa[1]*, Akihiro Ida[2], Tetsuya Hoshino[3], Ryosuke Ando[1]

## Summary

Large-scale earthquake sequence simulations using the boundary element method (BEM) incur extreme computational costs through multiplying a dense matrix with a slip rate vector. Hierarchical matrices (H-matrices) have often been used to accelerate this multiplication. However, the complexity of the structures of the H-matrices and the communication costs between processors limit their scalability, and they therefore cannot be used efficiently in distributed memory computer systems. Lattice H-matrices have recently been proposed as a tool to improve the parallel scalability of H-matrices. In this study, we developed a method for earthquake sequence simulations applicable to 3D nonplanar faults with lattice H-matrices. We present a simulation example and verify the mesh convergence of our method for a 3D nonplanar thrust fault using rectangular and triangular elements. We also performed performance and scalability analyses of our code. Our simulations, using over $10^5$ degrees of freedom, demonstrated a parallel acceleration beyond $10^4$ MPI processors and a >10-fold acceleration over the best performance when the normal H-matrices are used. Using this code, we can perform unprecedented large-scale earthquake sequence simulations on geometrically complex faults with supercomputers. The software HBI is made an open-source and freely available.

[1] Department of Earth and Planetary Science, University of Tokyo, Tokyo, Japan
[2] Research Institute for Value-Added-Information Generation (VAiG), Japan Agency for Marine-Earth Science and Technology (JAMSTEC), Yokohama, Japan.
[3] Information Technology Center, University of Tokyo, Chiba, Japan.
* corresponding author: sozawa@eps.s.u-tokyo.ac.jp

## 1 Introduction

Physics-based numerical simulations are an important tool in studying underlying mechanisms of earthquake processes. Among them, earthquake sequence simulations using rate and state friction laws, originating from Tse & Rice (1986) and Rice (1993), are widespread. Compared with single dynamic rupture simulations, which is the other widespread discipline in physics-based simulations of earthquakes, earthquake sequence simulations do not require the assumption of initial conditions as they solve the movement of the fault during both coseismic and interseismic periods in a single numerical framework. Several researchers now use earthquake sequence simulations to understand how faults behave under various conditions and how different model ingredients (e.g., fault rheology) influence an earthquake sequence (Erickson et al., 2020).

Among various computational methods, the boundary element method (BEM) is often used because of its ease in handling complex fault geometries (Hori et al., 2004; Ohtani et al., 2014; Qiu et al., 2016; Thompson & Meade, 2019; Yu et al., 2018). Another superiority of BEM is its relatively smaller computational cost. In fact, a recent comparison also highlights the computational cost of BEM is smaller than that of volume-discretized methods for a given grid spacing and accuracy (Jiang et al. 2022; Erickson et al. 2022).

Nevertheless, large-scale BEM simulations require huge computational costs. To increase the resolution of the simulation, characteristic element sizes often need to be reduced. In 3D simulations (2D fault in 3D space), if the characteristic element size is reduced by a factor of 2, the increase in $N$ is a factor of 4. In the original BEM, the computational cost for each time step scales with $O(N^2)$, where $N$ is the number of discretized elements. This is because multiplications of a dense matrix and a vector (slip rate distribution) are necessary to evaluate the stress change on each element at every time step. Furthermore, the time step width must be small if we use small elements, which increases the repetition of matrix-vector multiplications. Thus, the computational cost increases rapidly with a decrease in the element size.

Several methods have reduced the complexity of $O(N^2)$ to $O(N\log N)$. The fast Fourier transform (FFT) method is often used for this purpose (Kato, 2003; Lapusta & Liu, 2009), but is limited to planar or relatively simple fault geometries due to the assumption of translational symmetry (Romanet and Ozawa, 2022; Barbot, 2021). The FFT also cannot process non-vertical faults. Hierarchical matrices (H-matrices) (Hackbusch, 1999) are an alternative that can be used for general fault geometries. Ohtani et al. (2011) showed a significant acceleration in earthquake cycle simulations with H-matrices, and it is now common to use H-matrices in BEM-based quasi-dynamic earthquake sequence simulations (Galvez et al., 2020; Heimisson, 2020; Hyodo et al., 2016; Ohtani et al., 2014; Ozawa & Ando, 2021; Romanet, 2017). For example, Hyodo et al. (2016) performed earthquake sequence simulations in the Nankai trough megathrust using ~300,000 elements. H-matrices have also recently been used in elastodynamic problems (Sato & Ando, 2021; Chaillat et al. 2017).

A few parallelized libraries of H-matrices have been developed to deal with large-scale computation. For example, H-lib pro (Bebendorf & Kriemann, 2005) is a software for parallel H-matrices on distributed memory systems. Additionally, there are several parallel H-matrices libraries on GPU, such as HiCMA (Keyes et al., 2020), hmglib (Zaspel, 2019), and HACApK on GPU (Hoshino et al. 2018). In the earthquake modeling community, hmmvp (Bradley, 2014) was developed for arbitrary-shaped faults composed of rectangular elements, which was later used in QDYN (Luo et al., 2017; Galvez et al. 2020), an open-source earthquake sequence code.

However, Ida et al. (2014) showed conventional H-matrices had weakness in parallel scalability, which is important in computations using supercomputers. Owing to the increase in communication costs and load imbalance, the parallel speed increase is generally less than the expectation from the ideal linear scalability. Ida et al. (2014) showed that the computational speed of an H-matrix-vector multiplication saturates <100 cores in the Poisson equation of the $N{\sim}100{,}000$ problem. Thus, we could not efficiently use a large number of cores in the H-matrices.

Recently, Ida (2018) proposed the lattice H-matrices. The lattice H-matrices contain convenient structures to construct an efficient communication pattern compared with the normal H-matrices while maintaining the $O(N\log N)$ memory compression. In addition, a relatively adequate load balance is maintained in the case of lattice H-matrices, even if a large number of MPI processes are used. This method reduces the load imbalance and communication cost between MPI processes and improves parallel scalability, and it has been applied to micromagnetic simulations (Ida et al., 2020). The implementation of lattice H-matrices is freely available as an open-source in the HACApK library.

In this paper, we present a state-of-the-art quasi-dynamic earthquake simulator using conventional and lattice H-matrices, which is applicable to arbitrarily shaped fault(s) embedded in half-space. The computational code HBI is open-source and freely available under the MIT license. The code has also been validated against a benchmark problem for a planar strike-slip fault as defined by the Simulation of Earthquakes and Aseismic Slip (SEAS) project (see Jiang et al. (2022)). The structure of the manuscript is as follows. In Section 2, we describe the method for BEM-based earthquake sequence simulations. In Section 3, normal and lattice H-matrices are described. In Section 4, we show the simulation results and parallel scalability. Section 5 discusses and concludes the study.

## 2 Method of Earthquake Sequence Simulations on 3D nonplanar faults

The basic structure of our method is similar to many previous earthquake sequence simulations. Using BEM (section 2.1), the shear and normal stress changes due to slip are obtained. Coupling them with the rate and state friction law for each element leads to three ordinary differential equations (ODEs) as shown in section 2.2. To solve the time evolution problem, we use the Runge-Kutta method with error-based control of the step width.

### 2.1 Boundary Element Method

We assume a homogeneous and isotropic medium. The medium satisfies the equilibrium equation and Hooke's law with Lame constants $\lambda$ and $\mu$,

$$\sigma_{ij,j} = 0, \qquad (1)$$

$$\sigma_{ij} = \lambda \epsilon_{kk} \delta_{ij} + 2\mu \epsilon_{ij,} \tag{2}$$

where $\sigma_{ij}$ is the stress tensor, $\epsilon_{ij}$ the strain tensor, $\delta_{ij}$ the Kronecker's delta, respectively, and subscripts $i$ and $j$ runs $x, y, z$. Subscript $,j$ indicates the partial derivative with respect to the spatial coordinate $x_i$. The Einstein summation convention is used.

We use the boundary element method (BEM) to compute the elastic interaction (M. Bonnet, 1999), in which the elliptic partial differential equations (1-2) are transformed into integral equations. The shear stress change $\Delta\tau$ and normal stress change $\Delta\sigma$ are represented as the integral of the kernel function multiplied by the slip distribution on the fault surface (displacement discontinuity):

$$\Delta\tau(\boldsymbol{x}) = \int K_{shear}(\boldsymbol{x}, \boldsymbol{\xi}) \Delta u(\boldsymbol{\xi}) dS(\boldsymbol{\xi}), \tag{3}$$

$$\Delta\sigma(\boldsymbol{x}) = \int K_{normal}(\boldsymbol{x}, \boldsymbol{\xi}) \Delta u(\boldsymbol{\xi}) dS(\boldsymbol{\xi}), \tag{4}$$

where $K_{shear}$ and $K_{normal}$ are elastostatic integration kernels derived from the Green's functions (e.g., Segall, 2010), and $\Delta u$ is the slip distribution.

To numerically evaluate equations (3-4), we divide the fault surface into $N$ elements and denoted the index set as $I = \{1, \dots, N\}$. The shapes of the elements are either rectangular or triangular. In a discretized form using step functions as the base functions, the stress changes on the $i$-th element are represented as:

$$\Delta\tau_i = \sum_{j}^{N} A_{ij} D_j \,, \tag{5}$$

$$\Delta\sigma_i = \sum_{j}^{N} B_{ij} D_j \,. \tag{6}$$

where $D \in \mathbb{R}^N$ is the slip vector and $A$ and $B \in \mathbb{R}^{N \times N}$ are dense stiffness matrices. The entries of $A$ and $B$ are calculated using the half-space solutions of Nikkhoo & Walter (2015) and Okada (1992) for uniform slip (i.e., piecewise-constant interpolation) in triangular and rectangular elements, respectively. The evaluation point of the stress component is the center of each element. Triangular unstructured elements have more flexibility in fault geometry than rectangular elements. Note that Okada's solution also has a limitation that two parallel sides of an element must be horizontal. However,

Barall & Tullis (2016) found that rectangles outperform triangles in terms of the accuracy of the stress value. We will also compare the performance of triangular and rectangular meshes in later sections.

Notably, the normal stress change has often been neglected in several previous earthquake sequence simulations, unlike in single-event dynamic rupture simulations. Normal stress changes originate from broken symmetries such as nonplanar faults, free surfaces, and material heterogeneities.

## 2.2 Governing Equations

The boundary condition of each element is governed by the regularized rate and state friction law. Following Rice et al. (2001), the shear and normal tractions at each element are related as follows:

$$\frac{\tau_i}{\sigma_i} = a\mathrm{arcsinh}\left(\frac{V_i}{2V_0}e^{-\phi_i}\right), \tag{7}$$

where $V_i(t) = \frac{dD_i}{dt}$ is the slip rate, $\phi_i(t)$ is the state variable, $a$ is the coefficient of the direct effect, and $V_0$ is the reference slip rate.

The evolution law for the state variable is given by the aging law (Dieterich, 1979; Ruina, 1983):

$$\frac{d\phi_i}{dt} = \frac{b}{d_c}\left[V_0 \exp\left(\frac{f_0 - \phi_i}{b}\right) - V_i\right], \tag{8}$$

where $f_0$ is the reference friction coefficient, $b$ is the coefficient of the evolution effect, and $d_c$ is the characteristic slip distance. Using the stiffness matrices $A$ and $B$ calculated in the previous section, the shear and normal stress changes are given as follows:

$$\frac{d\tau_i}{dt} = \sum_j^N A_{ij}V_j + \dot{\tau}_i - \frac{\mu}{2c_s}\frac{dV_i}{dt}, \tag{9}$$

$$\frac{d\sigma_i}{dt} = \sum_j^N B_{ij}V_j + \dot{\sigma}_i, \tag{10}$$

where $\mu$ is the rigidity, $c_s$ is the S wave speed, $\dot{\tau}_i$ and $\dot{\sigma}_i$ are the tectonic loading rates for shear and normal stresses on the $i$-th element, respectively. The first terms in both equations (9-10) represent the stress rates caused by slip (time derivative of equations (5-6)). The third term for the shear stress is radiation damping, which is an approximation of inertia (Rice, 1993). Earthquake sequence simulations using this approximation are "quasi-dynamic," and the effect of this approximation has been explored in previous studies (e.g., Lapusta & Liu, 2009).

We eliminate $dV_i/dt$ from equation (9) using the total derivative of $V$:

$$\frac{dV_i}{dt} = \frac{\partial V_i}{\partial \tau_i}\frac{d\tau_i}{dt} + \frac{\partial V_i}{\partial \sigma_i}\frac{d\sigma_i}{dt} + \frac{\partial V_i}{\partial \phi_i}\frac{d\phi_i}{dt}, \tag{11}$$

so that:

$$\frac{d\tau_i}{dt} = \left(1 + \frac{\mu}{2c_s}\frac{\partial V_i}{\partial \tau_i}\right)^{-1}\left[\sum_j^N S_{ij}V_j + \dot{\tau}_i - \frac{\mu}{2c_s}\left(\frac{\partial V_i}{\partial \sigma_i}\frac{d\sigma_i}{dt_i} + \frac{\partial V_i}{\partial \phi_i}\frac{d\phi_i}{dt}\right)\right], \tag{12}$$

where the partial derivatives are (from equation (7)):

$$\frac{\partial V_i}{\partial \tau_i} = \frac{2V_0}{a\sigma_i}e^{-\phi_i}\cosh\left(\frac{\tau_i}{a\sigma_i}\right), \tag{13$a$}$$

$$\frac{\partial V_i}{\partial \sigma_i} = -\frac{2V_0\tau_i}{a\sigma_i^2}e^{-\phi_i}\cosh\left(\frac{\tau_i}{a\sigma_i}\right), \tag{13$b$}$$

$$\frac{\partial V_i}{\partial \phi_i} = -\frac{2V_0}{a}e^{-\phi_i}\sinh\left(\frac{\tau_i}{a\sigma_i}\right). \tag{13$c$}$$

Equations (13a-c) are substituted into equation (12). Equations (8), (10), and (12) form $3N$ ordinary differential equations (ODEs) $\frac{dy}{dt} = f(\boldsymbol{y})$ , where $\boldsymbol{y} = (\phi_1, \ldots, \phi_N, \tau_1, \ldots \tau_N, \sigma_1, \ldots, \sigma_N)$. We solve these equations using the Runge-Kutta method with adaptive time-stepping (Press et al., 2007). We compute $y(t + \Delta t_{try})$ with 5th order accuracy. If the maximum value of the relative difference between the 4th and 5th solution is larger than the allowance $\varepsilon_{RK}$, we retry the time integration as follows:

$$\Delta t_{new} = \max\left(\frac{\Delta t_{try}}{2}, \qquad 0.9\Delta t_{try}{\varepsilon_{RK}}^{-0.25}\right) \tag{14}$$

We choose $\varepsilon_{RK} = 10^{-4}$. If larger values are used, computational instability sometimes arises. If the error is below the threshold, we update the variables and calculate the next time step using the following formula:

$$\Delta t_{new} = \min\left(2\Delta t_{try}, \qquad 0.9\Delta t_{try}{\varepsilon_{RK}}^{-0.2}\right). \tag{15}$$

As a result, the time step is approximately inversely proportional to the maximum slip rate. This property results from the displacement in each time step having to be smaller than the characteristic state evolution distance. Lapusta et al. (2000) and many other studies explicitly adapted inverse-slip rate time-step widths based on stability analyses. The resultant $\Delta t$ weakly decreases with decreasing the element size if other parameters are identical. Finally, the slip is updated as follows:

$$D_i(t + \Delta t) = D_i(t) + \frac{\Delta t}{2}\left(V_i(t) + V_i(t + \Delta t)\right). \quad (16)$$

## 3 H-matrices

The most time-consuming step in solving the ODEs is the matrix-vector multiplication in equations (9-10) because it has an $O(N^2)$ complexity, whereas others display an $O(N)$ complexity (hereafter referred to as the $O(N)$ part). We aim to reduce the complexity of the matrix-vector multiplications to $O(N\log N)$ using conventional and lattice H-matrices. Our method is the same as Ida et al. (2014) and Ida (2018), except for the connection between the H-matrix-vector multiplication (HMVM) and the $O(N)$ part, which is specific to earthquake sequence simulations, In the implementation, we use the open-source library HACApK for the construction of H-matrices and HMVM. We will evaluate the performance of the normal H-matrices (section 3.1) and lattice H-matrices (section 3.2) in section 4.5.

### 3.1 H-matrices

H-matrices are an efficient method to compress the memory of the dense matrix derived from the integral operator (Borm et al., 2006; Hackbusch, 1999). The 3D elastostatic kernel ($K_{shear}(\boldsymbol{x}, \boldsymbol{\xi})$ and $K_{normal}(\boldsymbol{x}, \boldsymbol{\xi})$) exhibits $|\boldsymbol{x} - \boldsymbol{\xi}|^{-3}$ decay, and this kernel function can locally degenerate for a distant source and receiver points ($K(\boldsymbol{x}, \boldsymbol{\xi}) \sim \sum_k g_k(\boldsymbol{x}) h_k(\boldsymbol{\xi})$). This allows for constructing H-matrices for dense matrices $A$ and $B$ in equations (5-6) for typical mesh geometries. We construct H-matrices by the procedure in sections 3.1.1 and 3.1.2 before the time integration. To advance the one-time integration, we perform six HMVM for shear and normal stress rates (section 3.1.3). The parallelization technique is described in section 3.1.4.

### 3.1.1 Block partitioning

The construction of a block structure of an H-matrix consists of the following steps (Borm et al., 2006). First, we construct a binary cluster tree for the set of triangular or rectangular elements using the $(x, y, z)$ coordinates of their centers. We denote a cluster set as $\Omega_i$ $(i = 1, \ldots, N_\Omega)$. We set the minimum cluster size to 15 (Ida, 2018). Then, we construct a partition structure of the matrix using the following admissibility condition:

$$min\left(diam(\Omega_i), diam(\Omega_j)\right) < \eta\ dist(\Omega_i, \Omega_j) \tag{17}$$

where *diam* is the diameter of the cluster and *dist* is the distance between the two clusters. This condition is derived from the ability of the kernel function to approximately degenerate for distant source and receiver points. Typically, we set the parameter $\eta = 2$, following Ohtani et al. (2011).

Figure 1b shows an example of an obtained partition structure of the H-matrices for a rectangular-shaped fault plane (Figure 1a) divided by rectangular meshes. We reorder the index of the elements $I = (1, \ldots, N)$ according to the structure of the cluster tree in the construction of the matrix structure. Blocks located at far-diagonal parts tend to be larger than those around diagonal parts because they correspond to the interactions of distant locations, and the admissibility condition is easy to satisfy (see equation (17)). The partition structure is further described regarding the parallelization later.

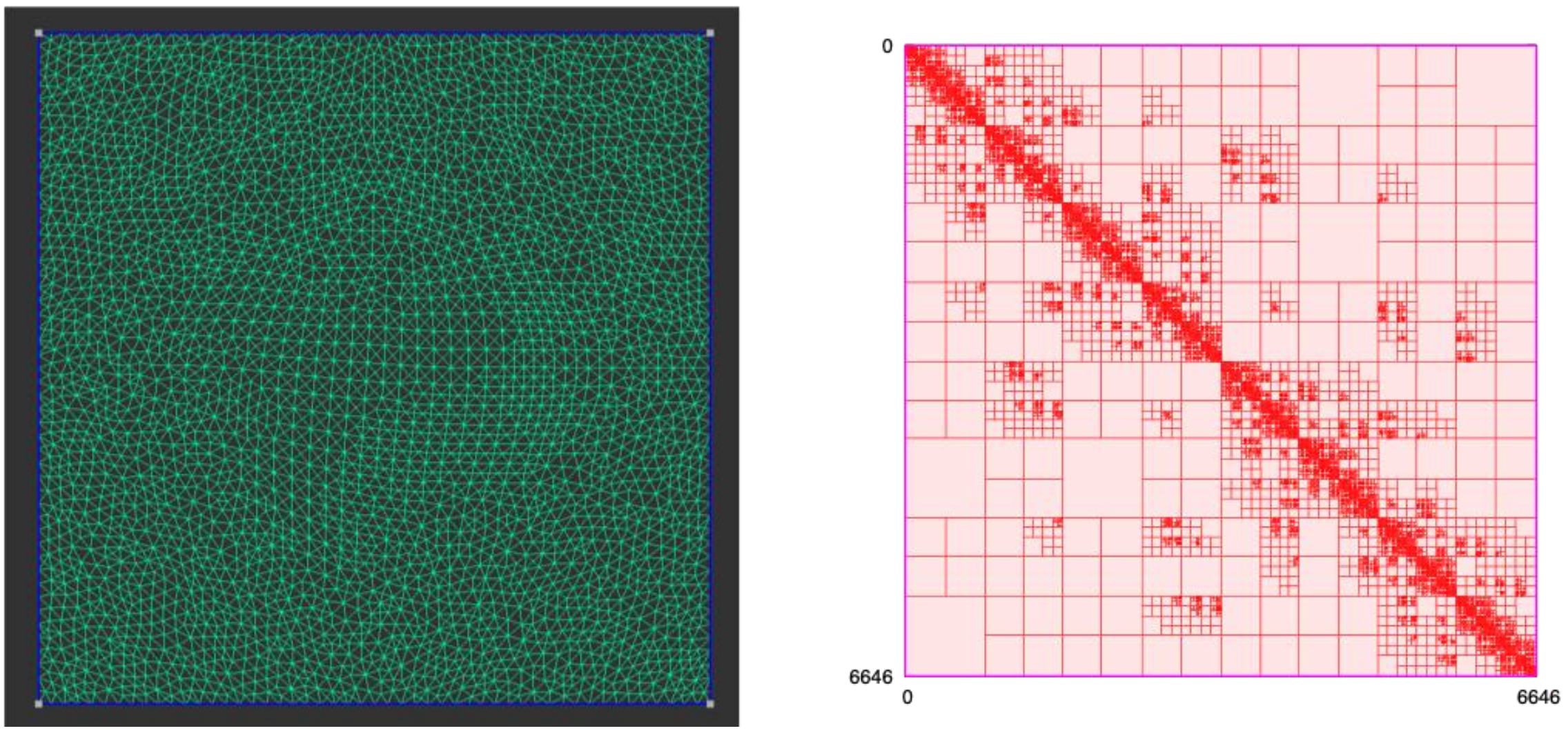

Figure 1: (left) A square-shaped fault using 6646 unstructured triangular elements created by Gmsh. (right) The corresponding block structures of H-matrices made by the admissibility condition $\eta = 2$.

### 3.1.2 Low-rank approximation (LRA)

Let $L, M \subset I$, and $A_{LM} \in \mathbb{R}^{L\times M}$ be a submatrix of $A \in \mathbb{R}^{N\times N}$. A submatrix $A_{LM}$ is compressed by a LRA if possible; otherwise, we use the dense matrix (full-rank matrix). A low-rank approximated submatrix $\tilde{A}_{LM}$ is represented as follows:

$$A_{ij} \approx \tilde{A}_{ij} = \sum_{k=1}^{r_{LM}} g_{ki} h_{kj}, \qquad (18)$$

where $g \in \mathbb{R}^{L\times r_{LM}}, h \in \mathbb{R}^{r_{LM}\times M}$, and $r_{LM}$ is the rank of the approximated matrix. As the LRA, we apply the adaptive cross approximation (ACA) (Bebendorf, 2000) whose computational complexity is O( $Lr_{LM}$ ) when $L \geq M$. Although the singular value decomposition is the optimal method for LRA, its complexity is O($LM^2$), which is too computationally expensive because we suppose $M \gg r_{LM}$. The rank $r_{LM}$ is controlled by the error tolerance, $\varepsilon_{ACA}$;

$$\frac{\|A - \tilde{A}\|_F}{\|A\|_F} < \varepsilon_{ACA}, \qquad (19)$$

where $\|\cdot\|_F$ denotes the Frobenius norm. Note that this condition is not rigorously achieved in ACA because we do not compute all entries of the matrix. Instead, in ACA we increase the rank one-by-one and stop the iteration once the difference becomes smaller than $\varepsilon_{ACA}$. We will evaluate the effect of the value of $\varepsilon_{ACA}$ in numerical experiments. We also use the method proposed by Ida et al. (2015) to prevent the H-matrix from having an excessively large rank.

### 3.1.3 H-matrix and vector multiplication (HMVM)

A matrix-vector multiplication $AV$ is performed submatrix-wise. For full submatrices $A_{LM}$ , the arithmetic is the normal matrix-vector multiplication. For low-rank approximated submatrices $\tilde{A}_{LM}$, we perform the arithmetic as

$$\sum_{j}^{L} A_{ij} V_j \approx \sum_{j}^{L} \sum_{k}^{r_{LM}} g_{ki} h_{kj} V_j = \sum_{k}^{r_{LM}} g_{ki} \left( \sum_{j}^{L} h_{kj} V_j \right). \tag{20}$$

The original computation using dense matrices requires $O(LM)$, while it becomes $O((L+M)r_{LM})$ in an HMVM. If the rank $r_{LM}$ is much smaller than $min(L, M)$, then the number of operations is significantly reduced. The arithmetic of equation (20) results in a part of the vector, and the full vector is obtained by taking the summation of all the submatrix-wise HMVMs.

### 3.1.4 Parallel Earthquake Sequence Simulation using H-matrices

Our earthquake sequence simulation code is parallelized using a message passing interface (MPI). Submatrices on the H-matrix are assigned to MPI processes, and each MPI process contains a quasi-1D-sliced portion of the entire matrix (Figure 3a). This does not represent a complete 1D slice because a submatrix cannot be separated into multiple MPI processes. For the HMVM, each processor possesses a full slip rate vector, but the resultant stress rate vector comprises a part of the slip rate vector in general. To obtain the full stress-rate vector, each MPI process calls MPI_iSEND and MPI_iRECV by $N_p - 1$ times, where $N_p$ denotes the number of MPI processes. The complexity of the communication cost is $O(NN_p)$. The algorithm is described in detail in Ida et al. (2014).

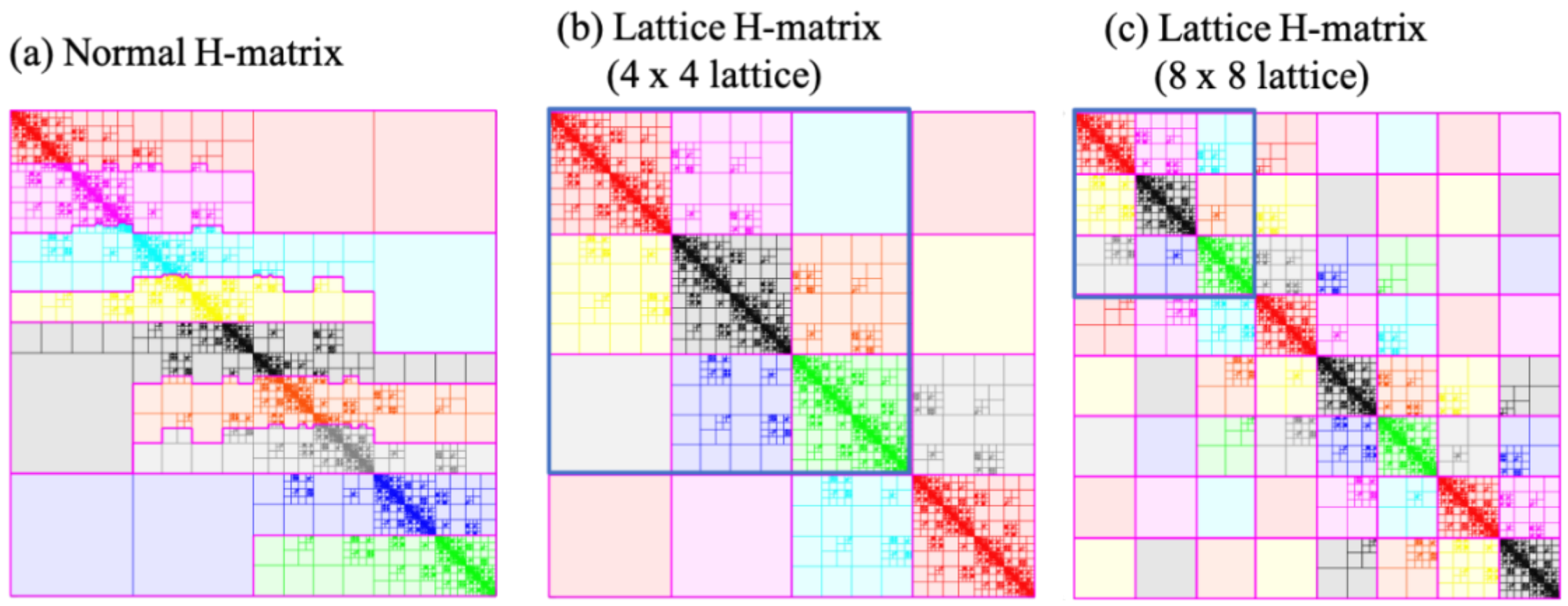

Figure 2: Comparison of the block structure and assignments to MPI processes of normal and lattice H matrices. Colors correspond to MPI processes $(N_p = 9)$. (a) is a normal H-

matrix. (b) and (c) represent lattice H-matrices using 3 x 3 process grid (shown in blue frames). (b) is a 4 x 4 lattice ($q = 1$) and (c) is an 8 x 8 lattice ($q = 2$).

For $O(N)$ part (element-wise computation), each MPI process is responsible for part of the vector. To construct the full-size vector required for the HMVM, MPI_Allgather is called before the HMVM. To perform a parallel computation of the $O(N)$ part, MPI_Scatter is called after the HMVM. As the number of MPI processes increases, the performance deteriorates owing to the MPI communication costs (both inside and outside the HMVM) in this method.

**3.2 Lattice H-matrices**

As explained above, earthquake sequence simulations using conventional H-matrices are not suitable for large-scale parallel computations because of the communication cost and load imbalance resulting from their extremely complex structure. To overcome this difficulty, Ida (2018) proposed lattice H-matrices. In this section, we describe the method for earthquake sequence simulations using lattice H-matrices. Hereafter, the H-matrices described in the previous section are referred to as normal H-matrices.

We first construct a cluster tree in the same way as the normal H-matrices, except that we truncate the depth $L$ of the cluster tree. We then construct a lattice structure using a truncated cluster tree. Then, an H-matrix is constructed for each lattice block in the same way as the normal H-matrices if it is admissible in terms of equation (17). The depth $L$ determines the number of lattice blocks (Figures 2b and 2c). As a result, the block structure of lattice H-matrices is simpler than normal H-matrices. LRA of each submatrix is the same as normal H-matrices.

We utilize the lattice structure for assigning MPI processes. This is achieved by introducing a process grid that has $N_{pr}$ rows and $N_{pl}$ columns ($N_{pr} \times N_{pl} = N_p$). We 2D-cyclically array this process grid on the lattice blocks, which means that each MPI process has discontinuous blocks of the matrix (Figures 2b and 2c). The number of lattice blocks is determined by the number of MPI processes, which ensures that $q$ process grids are repeated in rows and columns (Figures 2b and 2c). This condition gives $L = \lfloor \log_2(\sqrt{N_p}q) \rfloor$ because the binary tree is adopted. In a fixed $q$, as $N_p$ increases, each lattice becomes smaller. Thus, the entire matrix is divided into a larger number of submatrices, and the memory becomes larger compared with the normal H-matrices as confirmed later. However, in the procedure of HMVM using lattice H-matrices, we significantly reduce the communication traffic compared with the algorithm used in normal H-matrices. After the arithmetic of HMVM (equation (20)) assigned to each MPI process, diagonal MPI processes obtain part of the stress rate vector using MPI_Reduce along each row in the process grid, and then send it to other MPI processes in each column in the process grid using MPI_Bcast. This algorithm, which was first proposed by Ida et al. (2018) for block low-rank matrices, utilizes the lattice structure, and to perform this algorithm, the number of processors must be a squared number (Figure 3). Owing to the existence of the diagonal processes, only MPI communications between $\sqrt{N_p}$ processes are necessary, not all-to-all communication. It is also notable that only one MPI_Reduce and MPI_Bcast are called per each HMVM regardless of the number of MPI processes. Hence, the complexity of the communication costs for the HMVM using lattice H-matrices is $O(N)$, which is reduced from that of normal H-matrices $O(NN_p)$.

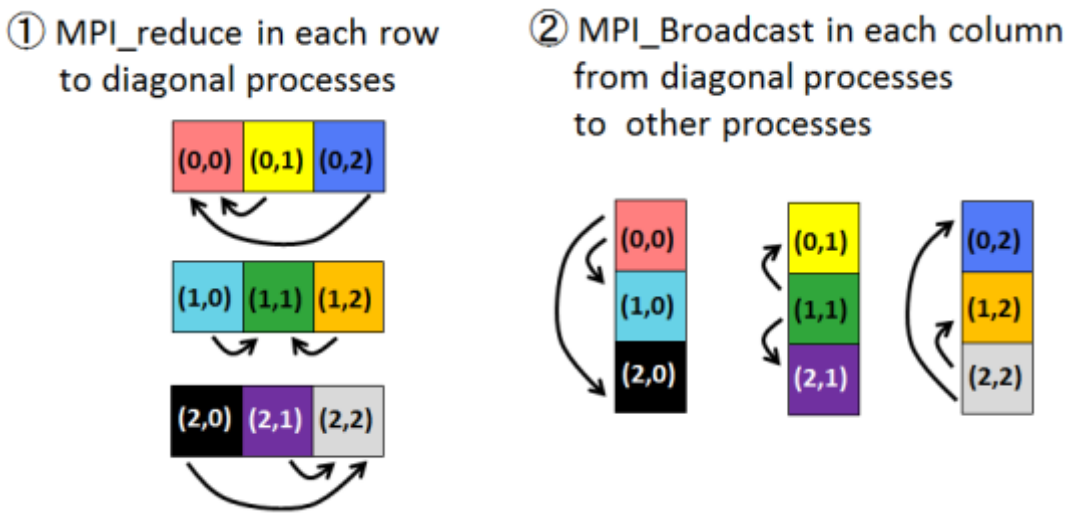

Figure 3: Schematic illustration of the algorithm of the MPI communication for HMVM (From Ida, 2018).

The HMVM in the lattice H-matrices requires only a part of the slip rate vector (size ~ $N/\sqrt{N_p}$) for each MPI process. In addition, each MPI process has identical indices of the resultant stress rate vector to the slip rate vector owing to the use of a squared number of processes. Because each MPI process is in charge of the same part of the vector for element-wise computation (the O(N) part) as the HMVM, unlike the normal H-matrix algorithm, MPI communication is not necessary before and after the HMVM. Note that this algorithm performs redundant computations for the O(N) part between $\sqrt{N_p}$ MPI processes. However, as confirmed later, HMVM comprises ~90% of the computational time in the case of $O(10^5)$ problems and a few tens of thousands of MPI processes, thus this redundant computation does not deteriorate the overall performance.

## 4 Numerical experiments

In this section, we perform numerical experiments using our code HBI, which implements the algorithm detailed in the previous sections. For convergence analysis, we use lattice H-matrices. For performance analysis we use both normal and lattice H-matrices.

### 4.1 Problem setting

A nonplanar fault is embedded in an elastic half-space, with elastic constants of $c_s = 3.464$ km/s, $c_p = 6$ km/s, and $\mu = 32.04$ GPa. The fault geometry is shown in Figure 4. The fault is 50 km in the along-strike length and 20 km in the along-dip length. The shallower (30° dip angle) and deeper (10° dip angle) parts are smoothly connected. The upper edge of the fault cut the free surface. We fix $b = 0.020$ and vary the *a*-*b* values, as shown in Figure 4 in color. We set $a/b = 0.75$ in the velocity-weakening zone. The characteristic slip distance $d_c$ is uniformly set to 0.02 m. The initial normal and shear tractions are uniformly set to 58 MPa and 100 MPa, respectively. For simplicity, we neglect the depth dependence of the initial shear and normal stresses.

The loading approach is the backslip method with a plate rate $V_{pl} = 10^{-9}$m/s for both the shear and normal stresses (e.g., Heimisson, 2020).

$$\dot{\tau}_i = -V_{pl} \sum_{j=1}^{N} A_{ij} \,, \tag{21}$$

$$\dot{\sigma}_i = -V_{pl} \sum_{j=1}^{N} B_{ij} \,, \tag{22}$$

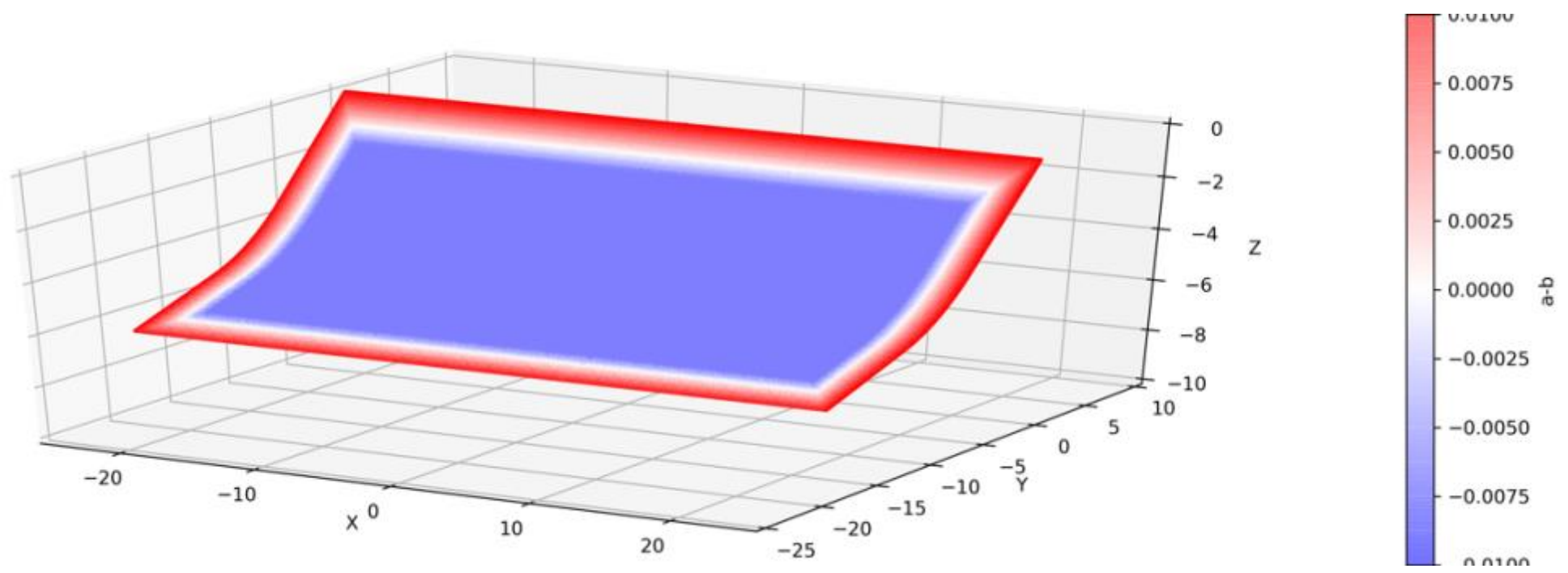

Figure 4: Fault geometry used in this study. The fault is 50 km in the along-strike length and 20 km in the along-dip length. There is a bend at dip = 10 km. The dip angle is 30° at the surface and 10° at the bottom. The color indicates the distribution of *a*-*b* values.

### 4.2 Simulation results and convergence test

Figure 5 shows the slip distribution of the first six events on the fault. Both partial ruptures and full ruptures occur in the velocity weakening zone of the fault. The interval of two partial ruptures ( ~100 days) is much smaller than the period of the earthquake cycle (~160 years), thus triggering of the latter rupture by the former rupture can be inferred. The cross section of cumulative slip is shown in Figure 6 and indicates that ruptures nucleate at the edge of the velocity weakening area and propagate toward the center of the velocity-weakening zone. It also shows the free surface produces significant coseismic slip despite its velocity-strengthening friction.

A previous study with a 2D planar fault demonstrated that, in this type of loading (back slip), the condition of the occurrence of the partial rupture is $W/h^* \gg 1$ (Cattania, 2019), where $W$ is the dimension of the velocity-weakening region and $h^*$ is given by $h^* = \frac{2\mu d_c}{\pi(b-a)\sigma}$. Otherwise, only full system size ruptures occur. We assume $W/h^*$>10 ($h^*$ ~2 km and $W$ >20 km), and the condition of partial ruptures is met. We believe that free surface effects and fault bends would also contribute to the earthquake sequence by modulating the elastic stress transfer, but a detailed discussion on the complexity mechanism is beyond the scope of this study.

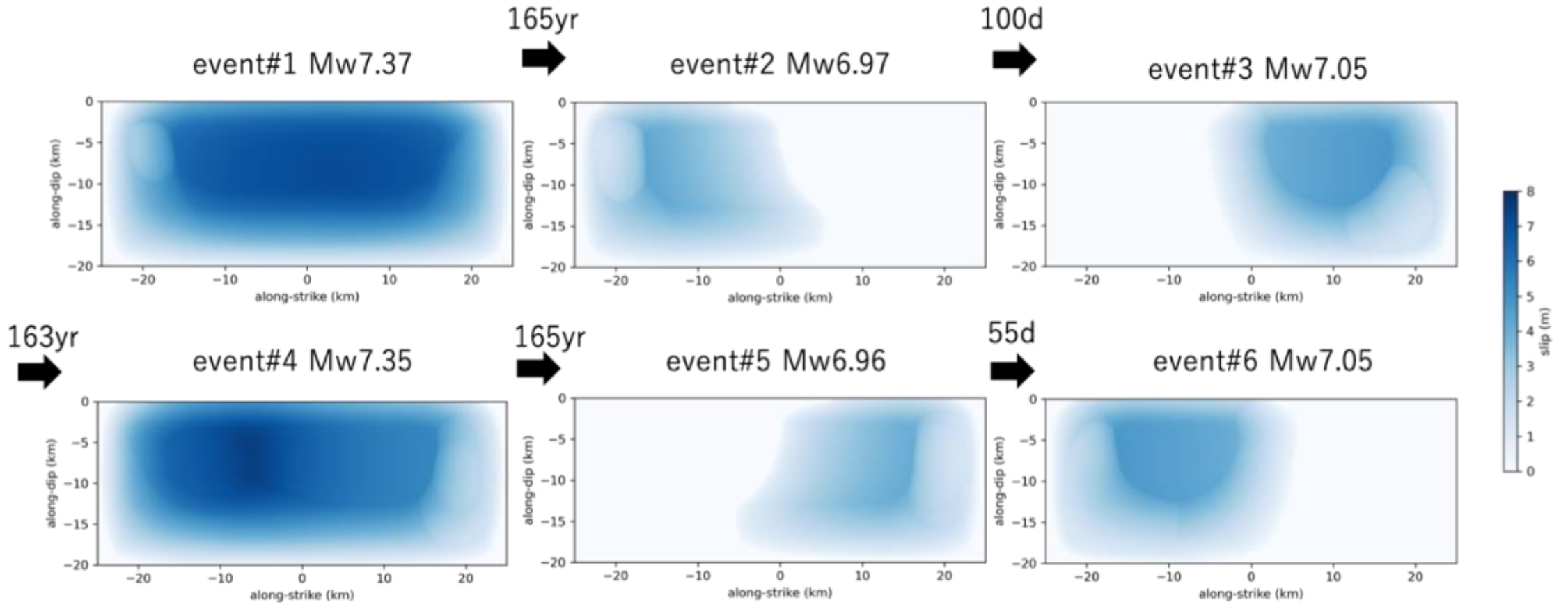

Figure 5: Slip distribution of the first 6 earthquakes. The result of 100,000 rectangular elements.

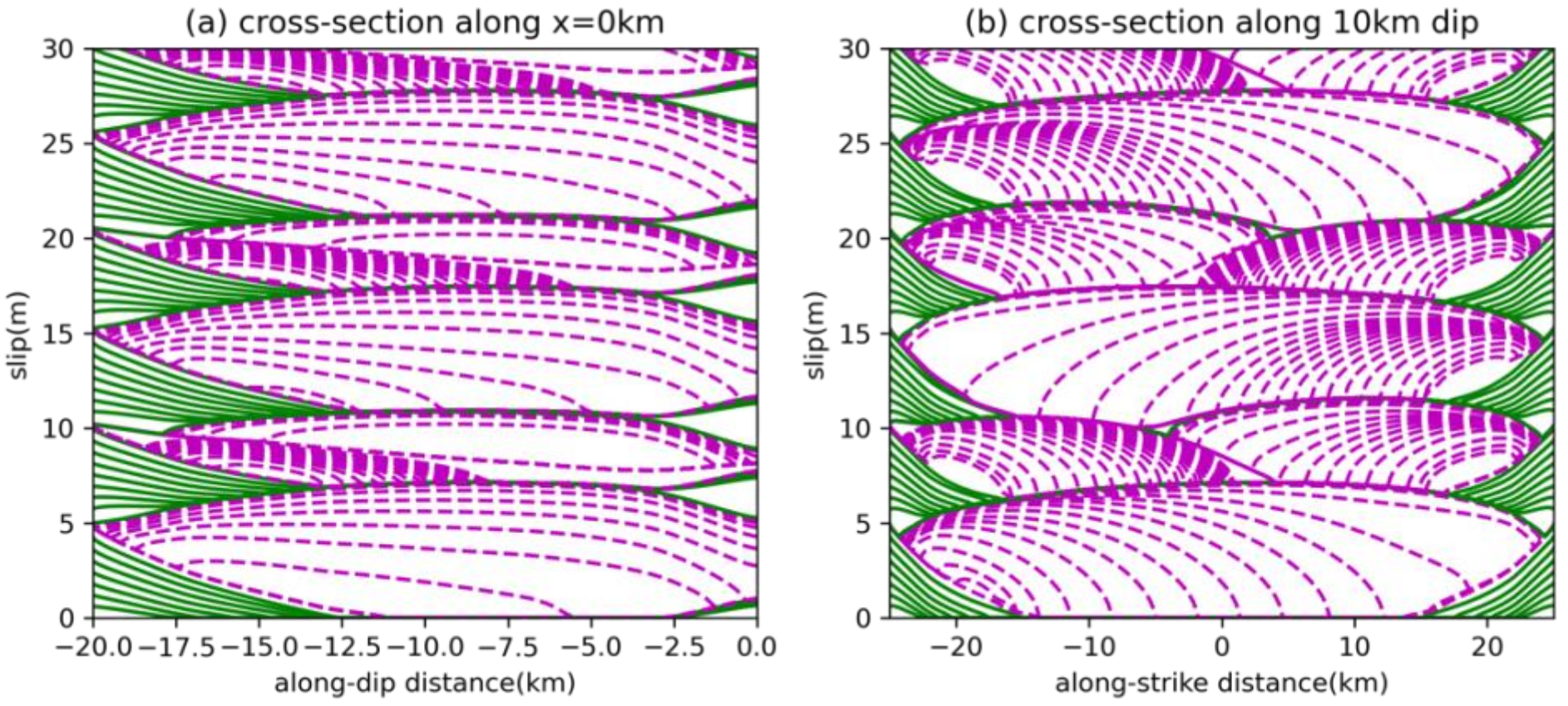

Figure 6. Cumulative slip at every 20 years during the interseismic period (green solid lines) and every 5 seconds during the coseismic period (purple dashed lines). (a) cross-section along $x = 0$ km. (b) cross-section along 10 km along dip. The result of 100,000 rectangular elements.

To attain the convergence of the numerical results, previous studies have shown that the following length scale must be resolved by at least a few elements (e.g., Rubin & Ampuero, 2005)

$$L_b = \frac{\mu d_c}{b\sigma}. \quad (23)$$

We perform a convergence test. Figure 7 shows the evolution of the mean friction coefficient using different mesh sizes for $\Delta s$. In the case of unstructured triangular elements, $\Delta s$ is defined as an input parameter of a free software Gmsh, which is used in creating unstructured meshes allowing the factor of 1.5 variability of the side lengths. Coarse meshes lead to different timing of ruptures, while finer meshes exhibit good agreements (Figure 7). It seems that $L_b/\Delta s > 3$ is enough for obtaining the same event history. However, the number of elements $N$ for triangular meshes is around twice as large as that of rectangular meshes for a given $\Delta s$. Thus, triangular elements need a larger number of elements than rectangular elements to attain an appropriate result.

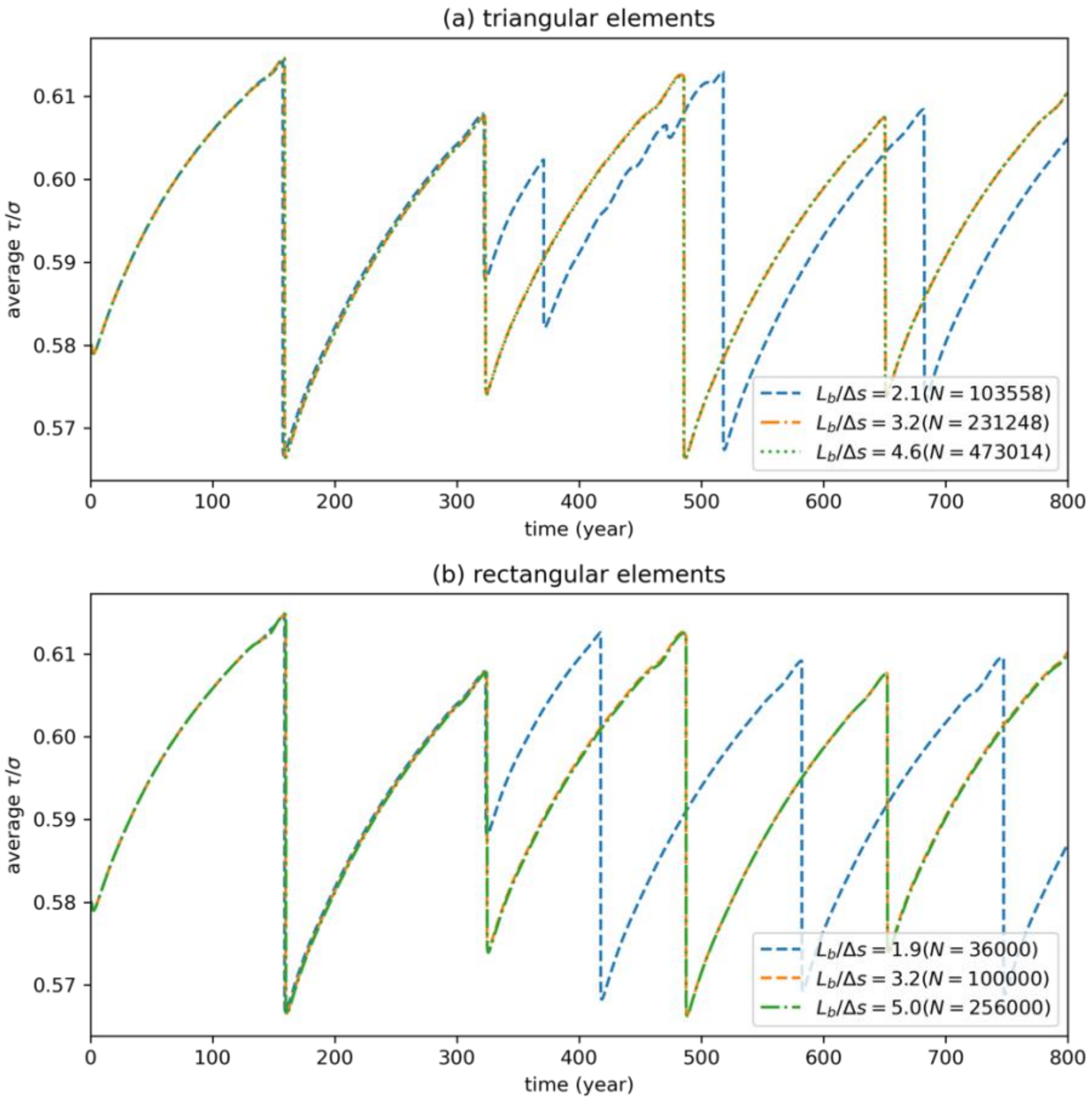

Figure 7: The time evolution of mean friction coefficient on the simulations using different element sizes. (a) rectangular elements and (b) triangular elements. The result of 100,000 rectangular elements.

### 4.3 Error due to low-rank approximation

We here examine the effect of approximation errors in LRA employed in our simulation. Figure 8 shows the results using different error tolerances of the H-matrices. All the cases show a good agreement, although the timing of the event has a discrepancy. For instance, the timing of the event that occurs around $t = 650$ years has 5 years difference between $\varepsilon_{ACA} = 10^{-2}$ and $10^{-3}$ and 0.5 year between $\varepsilon_{ACA} = 10^{-3}$ and $10^{-4}$. Ohtani et al. (2011) documented a larger discrepancy in the timing of the event than that observed here. We suspect that the nonuniform $d_c$ distribution (and thus nonuniform $L_b/\Delta s$ distribution) of Ohtani's model might be the cause of this discrepancy. Galvez et al. (2020) also reported that much smaller $\varepsilon_{ACA} = 10^{-8}$ is necessary to match the exact solution using highly heterogeneous $d_c$ distribution.

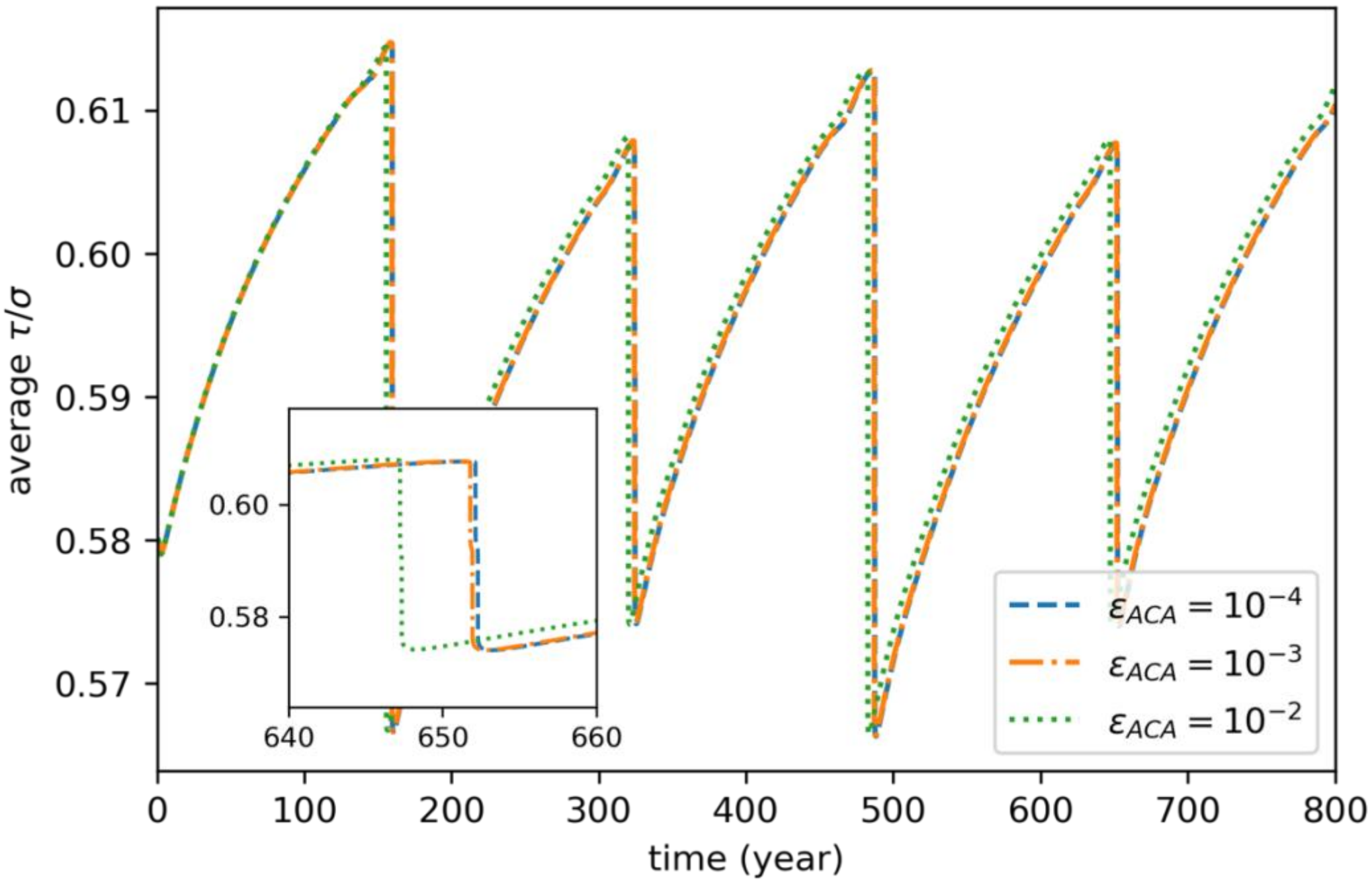

Figure 8: The time evolution of mean friction coefficient on the simulations using different error tolerance of the ACA in constructing a lattice H-matrix.

## 4.4 Memory Usage of H-matrices and Lattice H-matrices

As in Ohtani et al. (2011), we investigate the compression efficiencies of the normal and lattice H-matrices. Figure 9 shows the memory size of the normal H-matrices as a function of the number of elements. We fix the fault geometry and change the element size to vary the number of elements. We confirm a roughly $O(N\log N)$ dependence on the memory size for both the shear and normal stresses. For shear stresses and rectangular meshes, the compressibility against the original dense matrix is 8% for $N = 16{,}000$ and 0.7% for $N = 400{,}000$. Triangular meshes have larger memories than rectangular elements. This is presumably caused by the slightly non-uniform element sizes in unstructured meshes, which lowers the efficiency of the low-rank approximation of the submatrices.

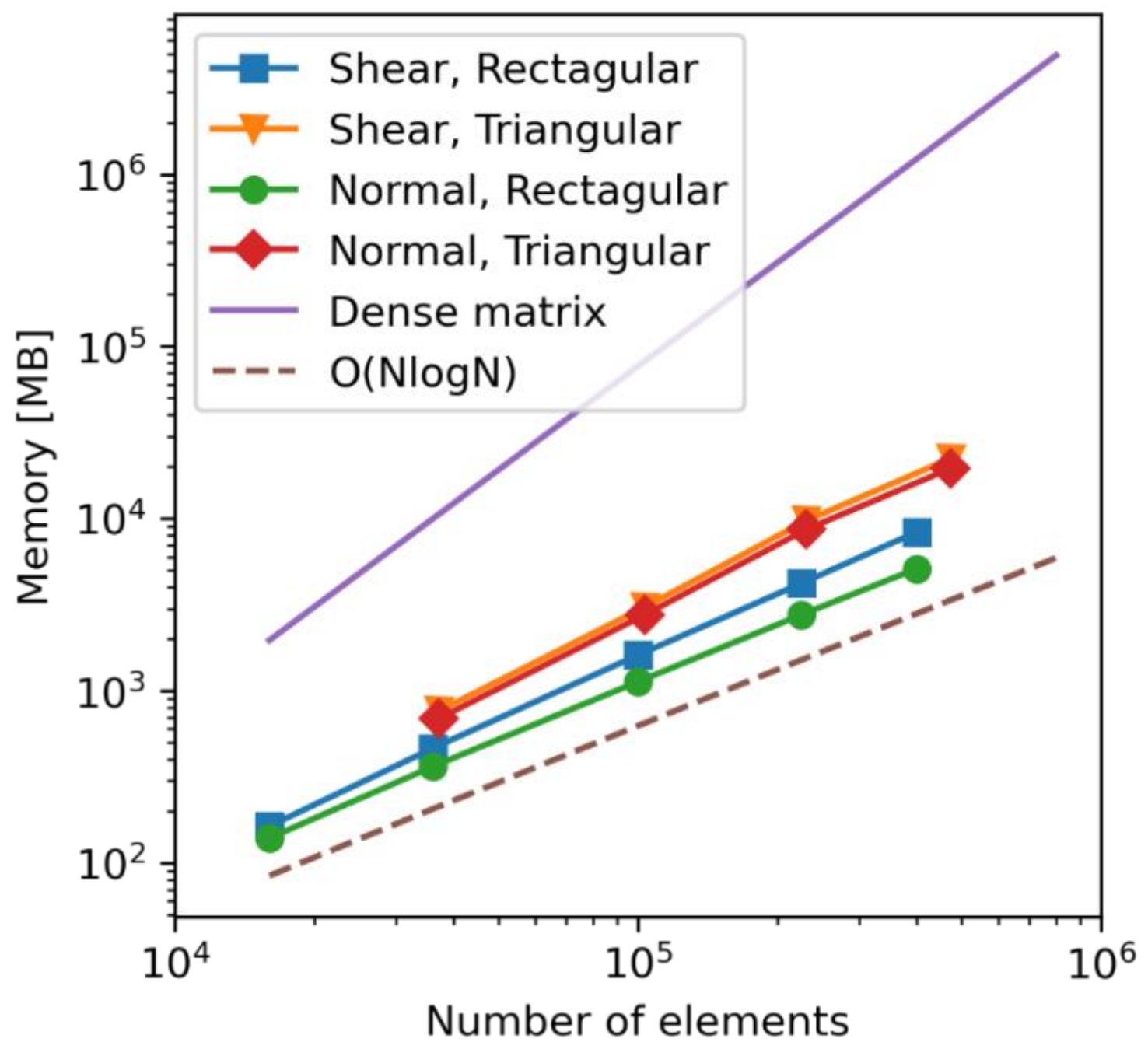

Figure 9: Memory sizes of the H matrix with respect to the number of elements. The memory size of the dense matrix ($O(N^2)$) and an $O(N\log N)$ slope is also shown as a reference.

Next, we measure the memory size of lattice H-matrices by varying the number of MPI processes (in principle, the memory size of the normal H-matrices does not depend on the number of MPI processes). We set $q = 4$ except for $N_p = 1$. As expected, the overall memory size of the lattice H-matrices increases with the number of MPI processes because of the smaller off-diagonal block sizes (Figure 10a). However, the maximum memory among the MPI processes is the bottleneck in the computation of HMVM, which is plotted in Figure 10b. For $N_p$<1,000, normal H-matrices are superior because the memory sizes of the diagonal MPI processes in the process grid tend to be large in lattice H-matrices. For $N_p$ >1,000, the lattice H-matrices show better load balance. The saturation of the maximum memory in normal H-matrices corresponds to the submatrix that has the largest memory.

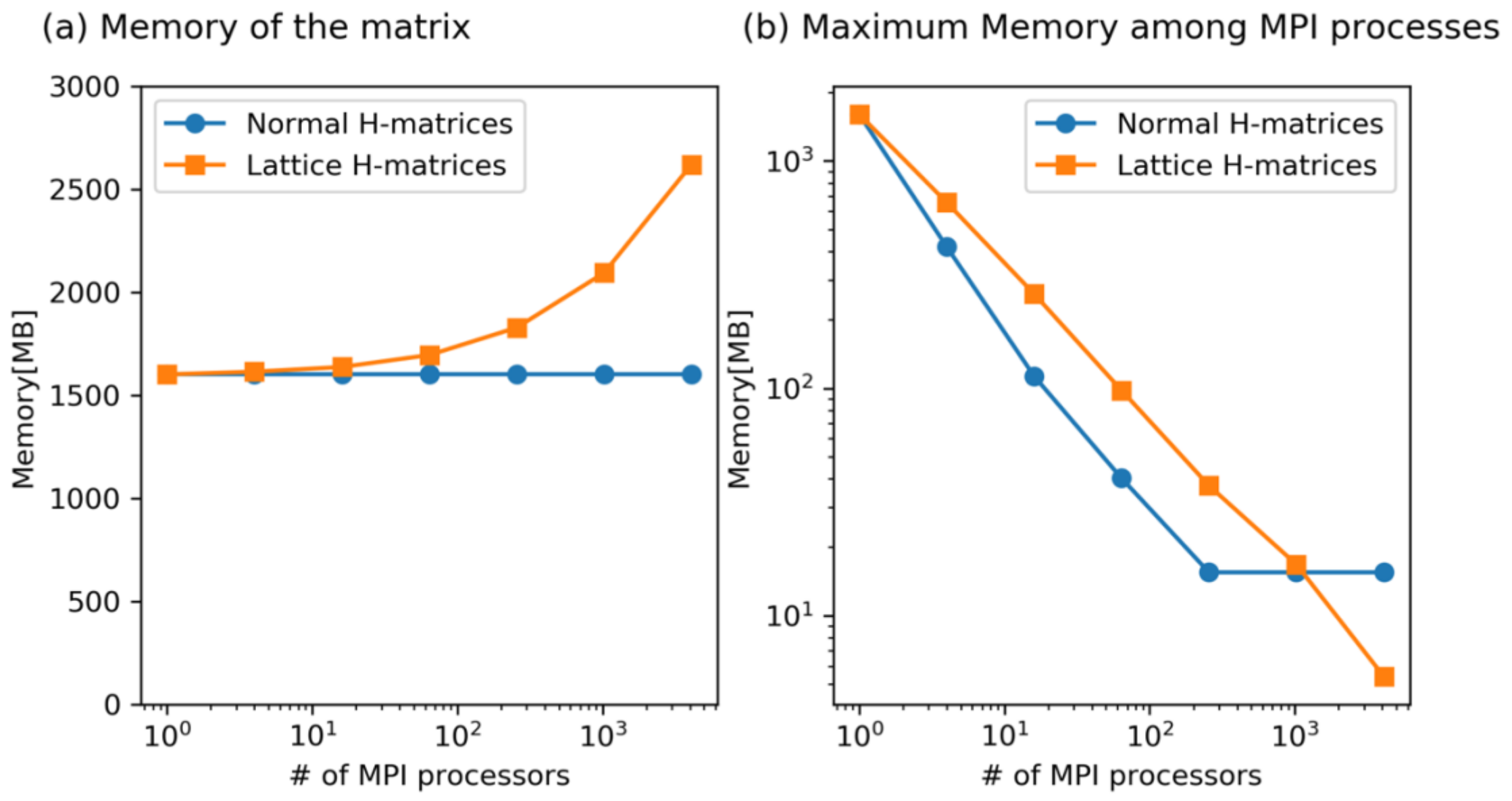

Figure 10: (a) Overall memory sizes of the normal and lattice H-matrices. (b) Maximum memory size among MPI processes. The case for shear stress and 100,000 rectangular elements.

### 4.5 Execution time and parallel scalability

Here we measure the execution time of the numerical simulations. All measurements were performed in Oakforest-PACS(OFP) at the University of Tokyo, which is equipped with an Intel® Xeon Phi ™ 7250 (68 cores, 1.4 GHz) and 96 GB(DDR4) memory in addition to 16 GB(MCDRAM) memory. The OFP system utilizes Intel® Omni-Path for the interconnect network, which has a link throughput of 100 Gbps. We used 64 cores per CPU node. We also used an Intel Fortran compiler with the -O3 optimization option and an Intel MPI Library. All results are flat MPI parallelization.

For lattice H-matrices, we measure the dependence of the number of elements on the execution time of 100 time steps using 100 and 900 MPI processes (Figure 11). As expected, we confirm a $O(NlogN)$ or slightly steeper increase in the execution time.

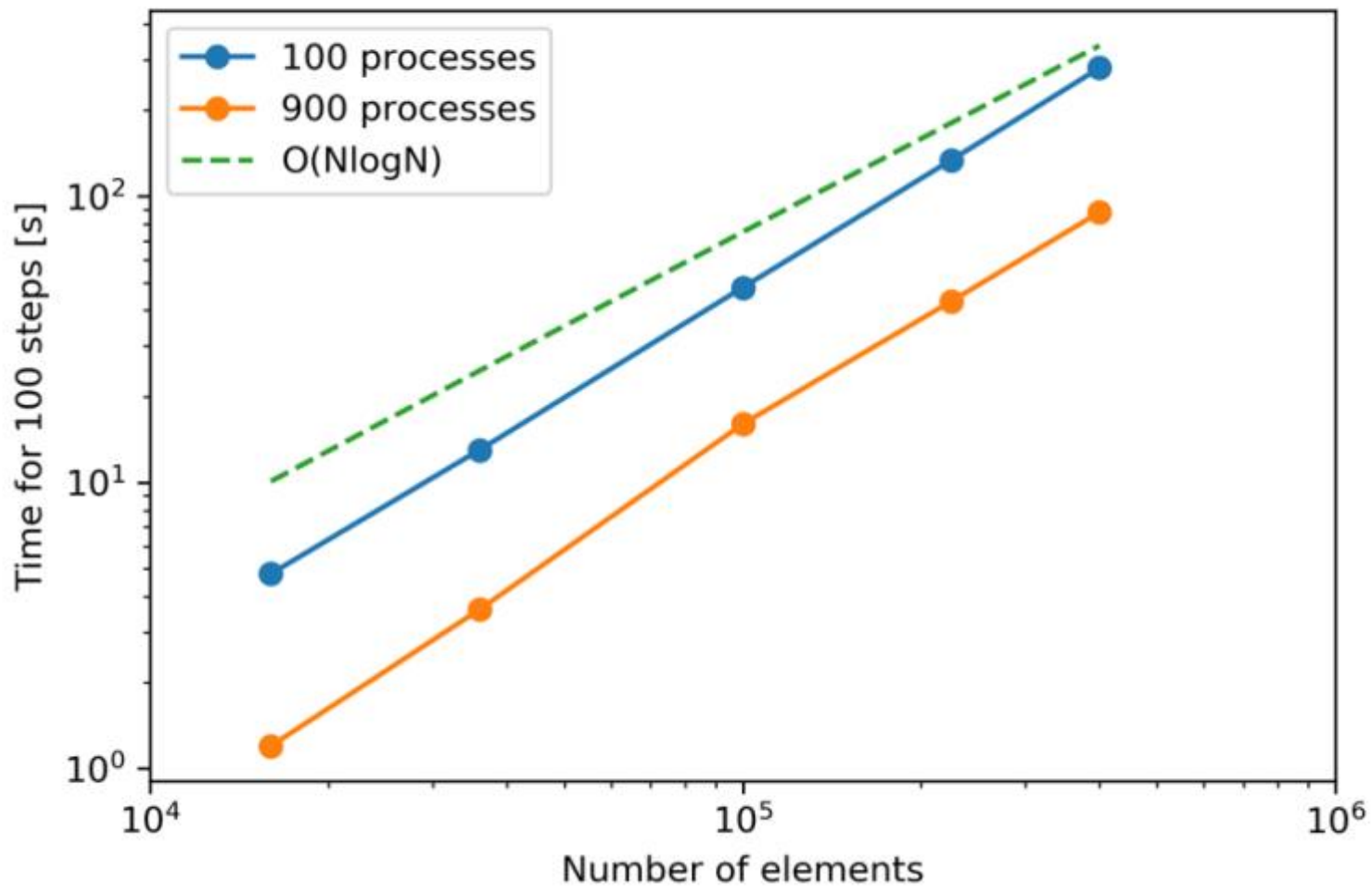

Figure 11: Number of elements vs. execution time of 100 time steps with the lattice H-matrices. $O(N\log N)$ curve is also shown as a reference.

Next, we measure parallel scalability (Figure 12). Our simulation using lattice H-matrices shows a consistent sublinear acceleration beyond 30,000 cores in the case of $N = 400{,}000$. We also show the result of the normal H-matrices, which exhibit an almost linear acceleration up to ~20 cores but rapidly saturate ~100 cores. The speed-down over 100 cores is caused by the increase in the communication cost, which is proportional to $N_p$.

By comparing the two methods, the normal H-matrices are faster by up to a few hundred MPI processes. The deceleration of lattice H-matrices from normal H-matrices occurred because the maximum memory for an MPI process is larger than that of normal H-matrices, as shown in the previous section (Figure 10b). The lattice H-matrices outperform normal H-matrices beyond a few 100s of cores owing to the reduction in the communication cost. We do not observe the saturation of the computation speed for lattice H-matrices, even with more than 10,000 cores. Figure 13 shows how a large fraction of the computation time is used in the HMVM in lattice H-matrices. The ratio of HMVM decreases with an increase in MPI processes, but it is always above 90%. From this figure, we expect a further acceleration in performance using additional processors.

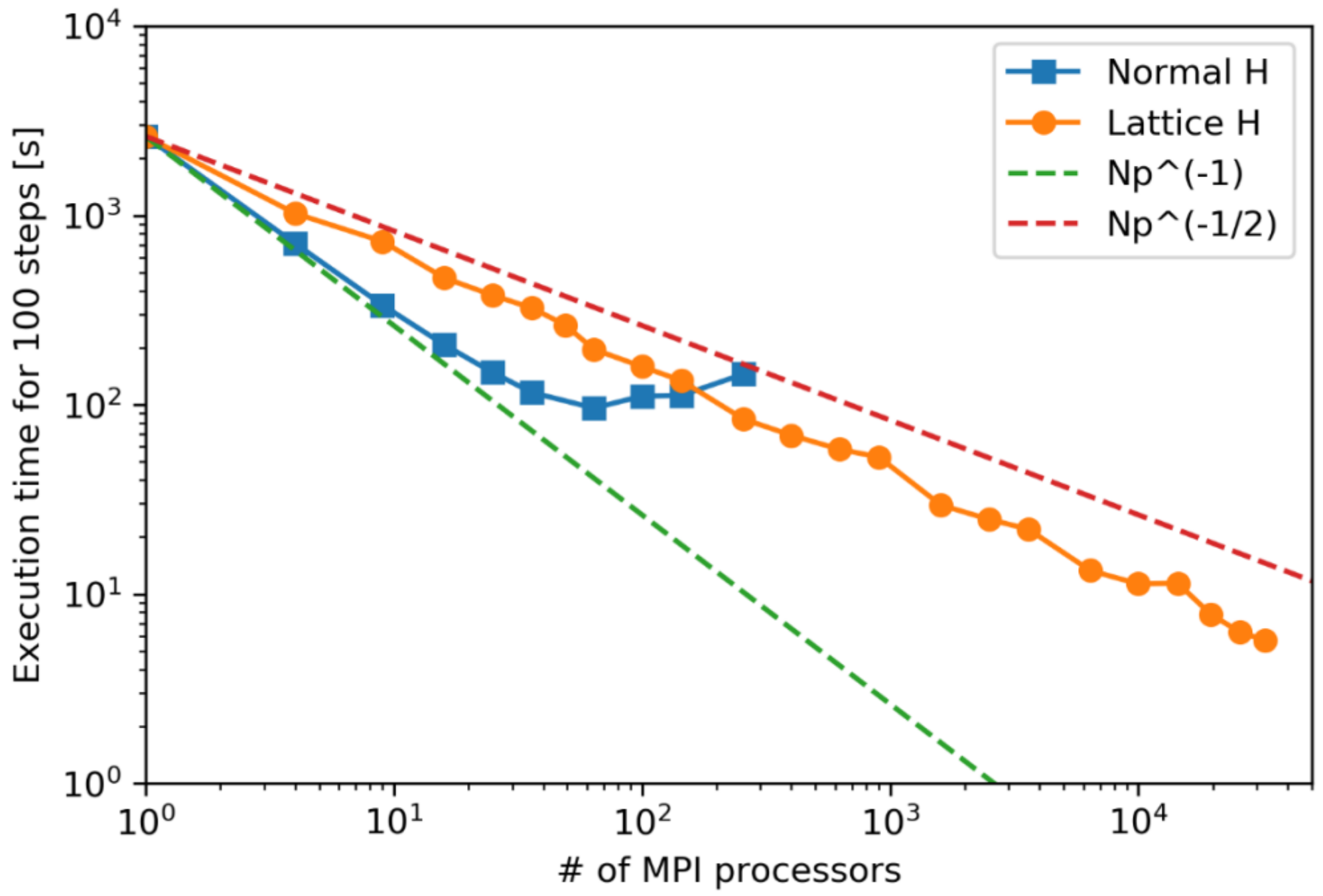

Figure 12: Parallel scalability when 100 time steps are performed (N=400,000 rectangular elements).

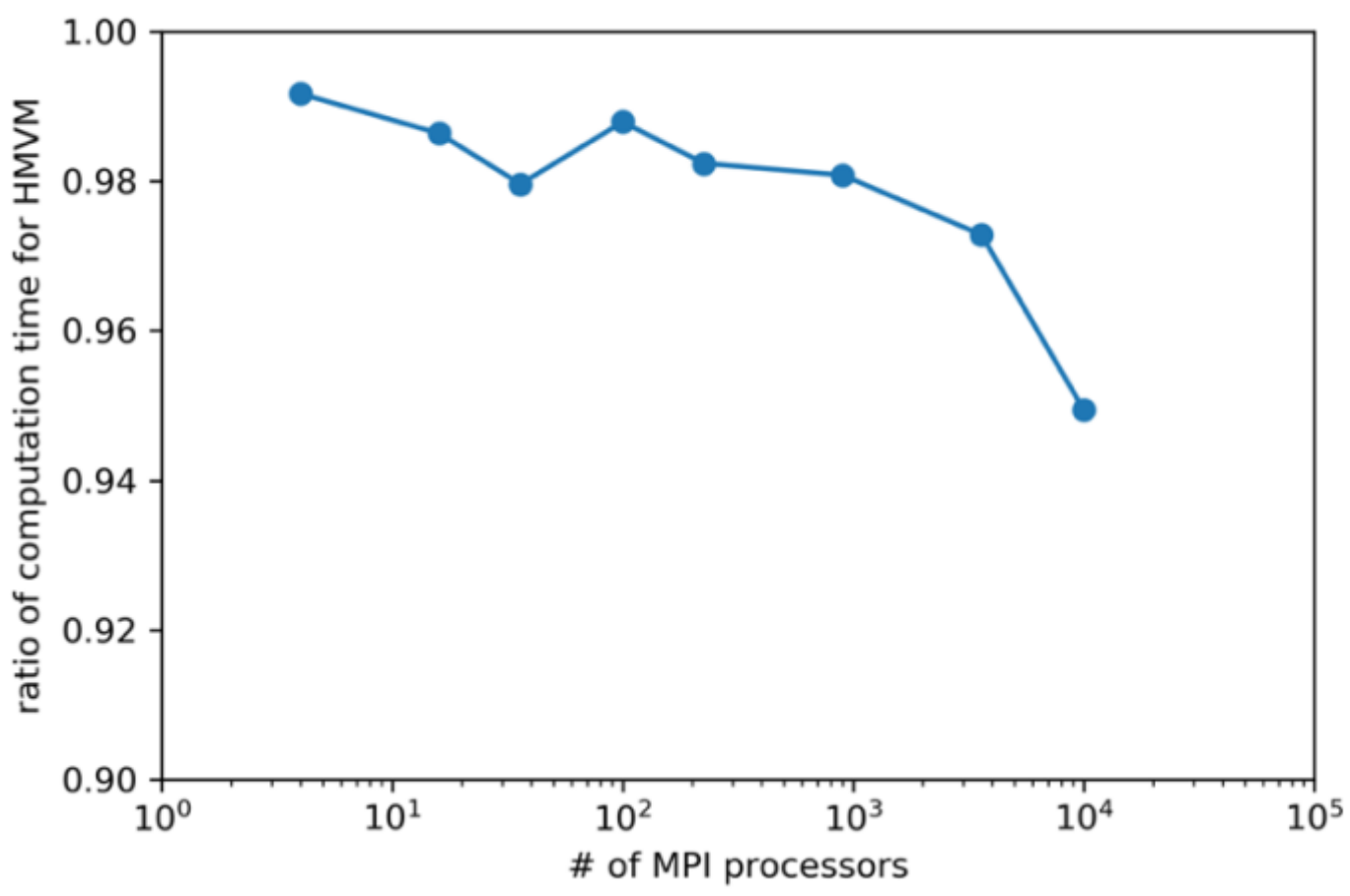

Figure 13: Ratio of calculation time of HMVM over a time step for N=400,000.

## 5 Discussion and Conclusions

In this study, we developed a method for earthquake sequence simulations with BEM using normal and lattice H-matrices. This method is highly flexible with fault geometry. Numerical experiments were conducted in a 3D nonplanar thrust fault to demonstrate the accuracy of our method in terms of convergence with decreasing mesh sizes. To our best knowledge, this work is the first comparison of triangular and rectangular elements in the context of earthquake sequence simulations. Our numerical simulation using a curved thrust fault showed complex patterns in earthquake sequences, which motivates us to conduct further studies focusing on earthquake physics . Our code can also be applied to natural fault systems worldwide and is potentially highly useful in physics-based earthquake hazard assessment.

In our numerical experiments, we confirmed $O(N\log N)$ complexity for the execution time for lattice H-matrices over 100,000 elements. Ohtani et al. (2011) observed a more rapid increase, and this was due to the increase of the rank of far off-diagonal blocks of the H-matrices. Hence, they had to set some artificial upper limit of the rank of submatrices to make large-scale simulations tractable. On the other hand, no matter how large $N$ was increased, any rapid increase of the rank was not observed in our simulations, even though our curved fault geometry is more complex than the planar fault model used by them. This might be due to the difference in the kernel: Ohtani et al. (2011) used Comninou & Dundurs (1975), while we used Nikkhoo & Walter (2015), which removed artifacts in previous solutions.

One question is whether this $O(N\log N)$ complexity of our result is maintained for further complex geometries, such as rough faults and/or fault networks (Ozawa & Ando, 2021). Additionally, inhomogeneous meshes can be used if the required resolution is not uniform due to spatial variation in friction and stress conditions. The use of inhomogenous meshes might change the compressibility of the dense matrices even for planar faults. Further studies are necessary to answer these questions.

We evaluated the parallel scalability of our simulation code using a supercomputer Oakforest-PACS. The lattice H-matrices overcame the high communication costs between MPI processes and enabled efficient computation using a large number of cores. The maximum computation speed for the lattice H-matrices was greater than ten times faster compared to the normal H-matrices. However, the lattice H-matrices were not as efficient as the normal H-matrices for a small number of cores. Thus, the lattice H-matrices should be used especially when a large number of CPUs is available.

Although we only performed flat-MPI simulations, we expect further acceleration using openMP and MPI hybrid parallelization. Hybrid parallelization is especially important for extremely large (N>1,000,000) problems, as only a few MPI processes can be used per CPU node because of memory limitations that cannot be distributed, such as the information of the coordinates.

**Acknowledgments**

This work is supported by JSPS KAKENHI [grant numbers 19J21676, 21H03447, 19K04031], the "Joint Usage/Research Center for Interdisciplinary Large-scale Information Infrastructures" and "High Performance Computing Infrastructure" in Japan (Project ID: jh210023-NAH), and MEXT under its Earthquake and Volcano Hazards Observation and Research Program. Numerical simulations were executed in Oakforest-PACS and Wisteria, supercomputer systems at University of Tokyo.

**Data availability**

All the simulation results are obtained using open-source software HBI (https://github.com/sozawa94/hbi).